\documentclass[12pt]{article}
\usepackage{osajnl}

\begin{document}

\title{Precise frequency measurements of atomic transitions using 
a Rb-stabilized resonator}
\author{Ayan Banerjee}
\author{ Dipankar Das}
\author{Vasant Natarajan}
\email{vasant@physics.iisc.ernet.in}
\affiliation{Department of Physics, Indian Institute of Science, 
Bangalore 560 012, INDIA}

\begin{abstract}
We demonstrate a technique for frequency measurements of atomic
transitions with a precision of 30 kHz. The frequency is measured using
a ring-cavity resonator whose length
is calibrated against a reference laser locked to the $D_2$ line of $^{87}$Rb, the 
frequency of which is known to 10 kHz. 
We have used this to measure the hyperfine 
structure in the $5P_{3/2}$ state of $^{85}$Rb. We obtain precise values
for the hyperfine constants $A=25.041(6)$ MHz and $B=26.013(25)$ MHz, and 
 a value of 77.992(20) MHz for the isotope shift in the $D_2$ line.
\end{abstract}
\maketitle

Precise measurements of atomic energy levels continue to play an important role in the 
development of physics. The energy levels of alkali atoms are particularly important 
because of their widespread use in laser-cooling experiments, ultra-cold collision studies, 
photoassociation spectroscopy, atomic tests of parity violation, and more recently in 
Bose-Einstein condensation. For example, precise measurement of the $D_1$ line in Cs 
\cite{URH99} combined with an atom-interferometric measurement of the photon recoil 
shift \cite{WYC93} could lead to a more precise determination of the fine-structure constant 
$\alpha$. In addition, hyperfine-structure and isotope-shift measurements in atomic lines 
can yield valuable information about the atomic wavefunction and the contribution from 
nuclear interactions. This is important when comparing theoretical calculations with 
experimental data in atomic tests of parity violation \cite{WBC97}.

In this Letter, we demonstrate a new technique for precisely measuring the 
frequencies of atomic transitions. The technique combines the advantages of using 
tunable diode lasers to access atomic transitions with the fact that the absolute 
frequency of the $D_2$ line in 
$^{87}$Rb has been measured with an accuracy of 10 kHz \cite{YSJ96}. A 
stabilized diode laser locked to this line is used as a frequency 
reference along with a ring-cavity resonator whose length is locked to the reference 
laser. For a given cavity length, an unknown laser locked to an atomic transition has a 
small frequency offset from the nearest cavity resonance. This offset is combined 
with the cavity mode number to obtain a precise value for the frequency of the 
unknown laser. We have used this method to make precise measurements of various 
hyperfine transitions in the $D_2$ line of $^{85}$Rb. This yields the hyperfine 
structure in the $5P_{3/2}$ state of $^{85}$Rb and the isotope shift in the $D_2$ 
line. We demonstrate a precision of 30 kHz, which is more than an order of 
magnitude improvement over the typical accuracy of other techniques such as 
level-crossing or optical double-resonance \cite{AIV77}. 
The only other technique with similar
precision is the frequency-comb method using mode-locked lasers \cite{URH99}. We believe 
our technique is uniquely suited for measuring hyperfine intervals and isotope shifts 
since the leading source of systematic errors cancels in such measurements. 

Similar measurements have been done previously using a stabilized HeNe laser as the 
frequency reference and a linear Fabry-Perot resonator \cite{BGR91}. Our use of a 
Rb-stabilized diode laser as the reference has the primary advantage that there are several 
known hyperfine transitions that can be used for locking the laser. This enables us to 
check systematic errors using different reference frequencies. The use of a ring-cavity 
resonator also has advantages over a linear resonator because the linear design
can cause feedback into the diode laser and destabilize it. By contrast, the 
ring cavity has a traveling wave inside and there is no possibility of 
feedback destabilization. Furthermore, the mode structure inside the cavity
is elliptical, which makes it easier to mode match the elliptical output of the diode laser. 

The schematic of the experiment is shown in Fig.\ 1. The experiment uses two 
frequency-stabilized diode 
lasers that are locked to atomic transitions using Doppler-free saturated-absorption 
spectroscopy. Laser 1 is the reference laser locked on the $D_2$ line of $^{87}$Rb. 
Laser 2 is the unknown laser, which in our case is locked on the $D_2$ line of 
$^{85}$Rb. The output of the two lasers is fed into the ring-cavity resonator. The 
cavity length is adjusted using a piezo-mounted mirror to bring it into resonance with 
the wavelength of Laser 1. The cavity is then locked to this length in a feedback loop. 
However, Laser 2 will generally be offset from the cavity resonance. This offset is 
accounted for by shifting the frequency of the laser using an acousto-optic modulator 
(AOM) before it enters the cavity. The error signal between the shifted frequency of 
Laser 2 and the cavity resonance is fed back to the AOM driver which locks the 
frequency of the AOM at the correct offset. The frequency is read using 
a frequency counter. The absolute frequency of Laser 1 is known with 10 kHz 
accuracy \cite{YSJ96}, therefore, once the cavity length (or mode number) is known, 
the frequency of Laser 2 is determined very precisely.

The diode lasers are standard external-cavity 
lasers stabilized using optical feedback from a piezo-mounted grating \cite{BRW01}. 
They are locked by modulating the injection current and using third-harmonic detection.
The ring cavity consists of two plane mirrors and two concave 
mirrors ($f=12.5$ cm) in a bow-tie arrangement. The mirrors are mounted on a copper 
plate that is temperature controlled to $\pm 0.01^{\circ}$C using a thermoelectric cooler. 
The cavity $Q$ 
is 35 and the free-spectral range (fsr) is 1.3 GHz (corresponding to a cavity length of 226.5 
mm). Thus the cavity modes (shown in Fig.\ 2) have a linewidth of 
about 37 MHz, which is similar to the 
linewidth of atomic transitions used in laser stabilization. The low $Q$ ensures robust 
locking that is quite insensitive to perturbations over a wide dynamic range.

The most important aspect of the measurement is to fix the mode number of the 
cavity uniquely. This is done in two steps: we first measure the cavity fsr
and then we use a known frequency for Laser 2. 
The fsr measurement proceeds as follows. We lock the cavity 
with the reference laser on the $F = 2 \rightarrow F' = (2,3)$ transition, as shown in 
Fig.\ 2a, and measure the AOM offset for a given transition of the unknown laser.
We then shift the reference laser to the $F = 1 \rightarrow F' = (1,2)$ transition, which is 
exactly 6622.887 MHz higher \cite{AIV77,YSJ96}. This shift causes the cavity mode number to 
increase by almost exactly 5, as shown in Fig.\ 2b. The cavity is locked to the new 
frequency and the AOM offset for the same transition of the unknown laser is measured. The 
difference in the two AOM offsets along with the change in the reference frequency 
gives exactly 5 times the cavity fsr. Using this method, the fsr is determined 
with a precision of 20 kHz. To determine the mode number, we use a diode 
laser stabilized on the Rb $D_1$ line as Laser 2 and determine its AOM offset. The 
frequency of this laser is already known to an accuracy of 0.4 MHz \cite{BGR91}. Thus,
there is a unique mode number that matches 
the resonance condition and the measured fsr. For example, the next nearest mode 
satisfying the resonance has an fsr differing by 350 kHz, or 17 times the error 
in the determination of the fsr. Similarly, if the mode number is changed
by 1, the inferred frequency of Laser 2 changes by 30 MHz,
which is 75 times larger than the error with
which the frequency is known.

We have used the technique to measure hyperfine structure in Rb. To test the 
reliability of the measurement and our error budget, we first measured hyperfine 
transitions in the $D_2$ line of $^{87}$Rb where the frequencies are already known to an 
accuracy of 10 kHz. For example, for the $F = 2 \rightarrow F' =(1,3)$ transition, 
we measure a frequency of 384 227 903.438(30) MHz while the value from Ref.\ 4 
is 384 227 903.407(10) MHz.
With these checks, we have measured hyperfine transitions in $^{85}$Rb. 
The results are listed in 
Table 1. Each value is an average over 50 individual measurements. 
The 5 measurements listed are not all 
independent because there are only 4 hyperfine levels 
in the $5P_{3/2}$ state. 
Measurements 2 and 4 couple to the same excited level from different ground levels, 
therefore their difference should be the ground hyperfine splitting of 3035.732 MHz 
\cite{AIV77}. The measured difference of 3035.682(45) MHz is a further check on our error 
budget. We have also repeated these measurements with the reference laser
on the $F = 1 \rightarrow F' = (1,2)$ transition, and verified that the results are consistent
within the quoted errors. Finally, the measurements were repeated over a period of several weeks to
check for any long-term variation.

The several internal consistency checks discussed above check one class of systematic errors, 
namely those that are wavelength independent. They could arise because of shifts in the 
lock point
of the lasers, peak pulling from residual Doppler profile, optical pumping effects, etc.
In addition, we have verified that systematic errors due to stray magnetic fields or collisional 
shifts in the Rb vapor cell are negligible. The magnetic field in the vicinity of the cell 
is less than 0.1 mT, and we have repeated the 
measurement with the vapor cell at different locations to check that the shift is negligible.
Collisional shifts in the vapor cells are estimated to be less than 10 kHz, and we 
have used different vapor cells to verify that these shifts are also negligible.

There is another class of systematic errors that is wavelength dependent. Since our cavity is 
comparing the {\it wavelengths} of the two lasers, we have to convert these to frequencies using 
the refractive index of air \cite{EDLa,EDLb}. Any error in the refractive index would reflect as a 
systematic shift in the measured frequency. The ratio of refractive indices at two 
wavelengths from Ref.\ 8 is expected to be correct only at the $10^{-9}$ level, 
therefore, in future, we plan to use an evacuated cavity to 
eliminate dispersion effects. Such errors can also arise 
from wavelength-dependent phase shifts in the cavity mirrors. The 
dielectric coatings have extremely flat response over $\sim$200 nm and we do not expect this 
to affect measurements at the 30 kHz level over a range of 100 nm.

It is important to note that this class of wavelength-dependent systematic errors does not 
affect the determination of {\it frequency differences} of the unknown laser, up to several 10s 
of GHz. This is the order of magnitude of frequency differences that arise when 
determining hyperfine structure or isotope shifts. Thus, in the present case, we are
confident that such systematic errors are negligible in the determination of the hyperfine 
intervals in $^{85}$Rb. Indeed, the absolute frequencies of the various transitions in 
Table 1 are also correct because both lasers are at the same wavelength of 780 nm and 
there is no need to correct for the dispersion of air.

We have used the data in Table 1 to obtain the hyperfine coupling constants in the 
$5P_{3/2}$ state of $^{85}$Rb. The measured intervals are fitted to the magnetic-dipole 
coupling constant $A$ and the electric-quadrupole coupling constant $B$, to yield 
values of $A=25.041(6)$ MHz and $B=26.013(25)$ MHz. In Fig.\ 3a, we show the 
good agreement between the measured intervals and the intervals calculated from the 
fitted constants. The $A$ and $B$ values are compared to the earlier values of 
Arimondo et al.\ \cite{AIV77} and Barwood et al.\ \cite{BGR91} in Fig.\ 3b. Our value of
$A$ is consistent with the previous values but has much smaller error. However,
the three values of $B$ are quite discrepant and have non-overlapping error bars, 
though our value overlaps with the recommended value of Arimondo et al.\ at 
the $2\sigma$ level. Finally, the calculated hyperfine shifts can be used to obtain
the hyperfine-free $D_2$ frequency, which in turn yields the isotope shift in this
line to be 77.992(20) MHz.

In conclusion, we have demonstrated a new technique for precisely measuring the 
frequency of atomic transitions. We have used this 
technique to measure hyperfine transitions in the $D_2$ line of $^{85}$Rb with a 
precision of 30 kHz. Since the reference and unknown lasers are diode lasers locked to atomic
transitions using similar techniques, the accuracy and stability of our measurements are both 
determined by how well we can lock the lasers to the line-center. In future, we plan 
to use an evacuated cavity where the leading source of systematic 
error due to dispersion of air will be 
overcome. This will enable us to measure transitions in other alkali atoms 
(such as Li, K, and Cs) where the transitions are similarly accessible using diode lasers, 
but the difference from the reference frequency is much larger. Another improvement 
is to mix both beams in an optical fiber 
for better mode-matching into the cavity. We believe the technique 
can also be extended to UV atomic lines by using a frequency-doubled IR laser. By measuring 
the IR frequency instead of the UV frequency, we can get similar accuracy for UV lines 
without the concomitant sources of systematic error.

This work was supported by the Board of Research in Nuclear 
Sciences (DAE), and the Department of Science and Technology, Government of India.


\newpage
\begin{table}
\caption{ 
The table lists the measured frequencies of various hyperfine 
transitions in the $D_2$ line of $^{85}$Rb. }
\begin{tabular}{lc}
\multicolumn{1}{c}{Measured transition} & Frequency (MHz) \\
\hline
1. $F=3 \rightarrow F'=(3,4)$ & 384 229 181.506(30) \\
2. $F=3 \rightarrow F'=(2,3)$ & 384 229 089.323(30) \\
3. $F=3 \rightarrow F'=(2,4)$ & 384 229 149.816(30) \\
4. $F=2 \rightarrow F'=(2,3)$ & 384 232 125.005(30) \\
5. $F=2 \rightarrow F'=(1,3)$ & 384 232 110.353(30) \\
\end{tabular}
\label{freqs}
\end{table}

\begin{figure}
\scalebox{0.7}{\includegraphics{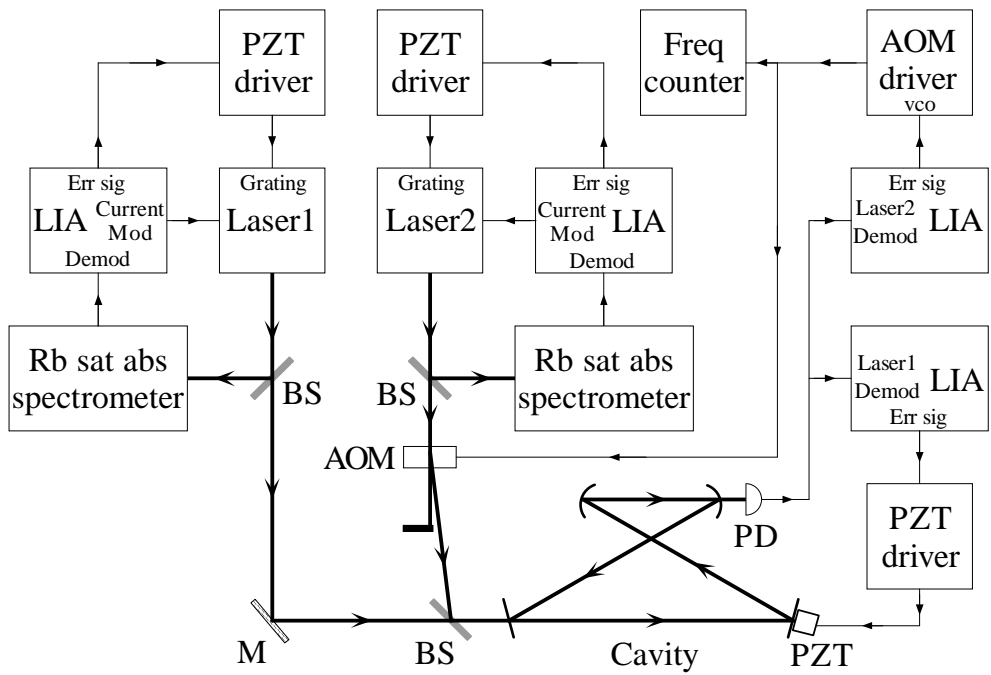}}
\caption{
Schematic of the experiment.}
\end{figure}

\begin{figure}
\scalebox{0.7}{\includegraphics{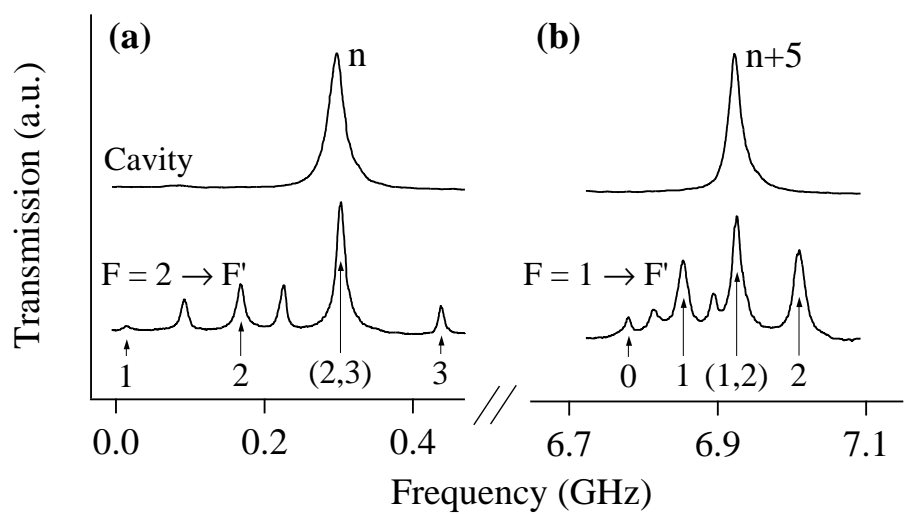}}
\caption{
The cavity fsr is determined using two lock points for the reference laser:
$F=2 \rightarrow F'=(2,3)$ shown in a), and $F=1 \rightarrow F'=(1,2)$ shown in b). The 
cavity mode number in b) increases by exactly 5.}
\end{figure}

\begin{figure}
\scalebox{0.7}{\includegraphics{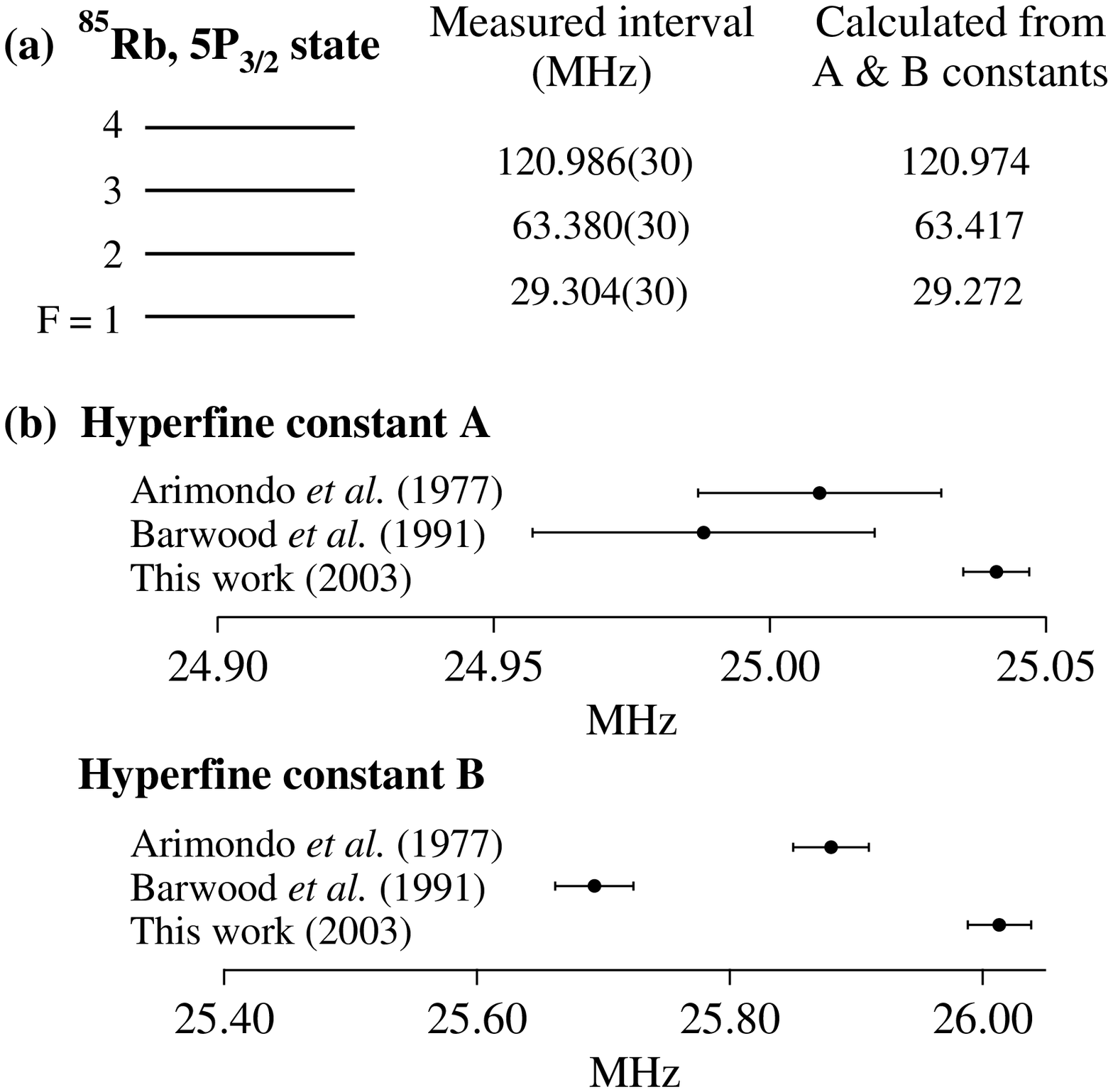}}
\caption{
Hyperfine-structure in the $5P_{3/2}$ state of $^{85}$Rb. In a), we compare the
measured and calculated values of the hyperfine intervals. In 
b), we compare our $A$ and $B$ values with the earlier values reported by Arimondo et 
al.\ \cite{AIV77} and Barwood et al.\ \cite{BGR91}. }
\end{figure}

\end{document}